\documentclass[11pt, oneside]{article}   	
\usepackage{graphicx}				
\usepackage{caption}
\usepackage{subcaption}
\usepackage[margin=1in]{geometry}
\usepackage{setspace}
\doublespace

\usepackage{fancyhdr}

\usepackage{amssymb}
\usepackage{amsmath}
\usepackage{hyperref}
\usepackage[usenames,dvipsnames]{color}
\usepackage{colortbl}

\title{Predictive Modeling of Opinion and Connectivity Dynamics in Social Networks}
\author{Ajay Saini $^{1}$ and Natasha Markuzon $^{2}$\thanks{Corresponding author: Dr. Natasha Markuzon (Email: nmarkuzon@draper.com, Phone: 617-258-2740, Fax: 509-693-7397), Draper Laboratory, 555 Technology Square Cambridge, MA 02139.} \\ $^{1}$Acton-Boxborough Regional High School, MIT PRIMES, ajays725@gmail.com \\ $^{2}$Draper Laboratory, nmarkuzon@draper.com}

\date{}							

\begin{document}

\maketitle

\thispagestyle{empty}
\clearpage

\thispagestyle{empty}
\begin{abstract}
Recent years saw an increased interest in modeling and understanding the mechanisms of opinion and innovation spread through human networks. Using analysis of real-world social data, researchers are able to gain a better understanding of the dynamics of social networks and subsequently model the changes in such networks over time. We developed a social network model that both utilizes an agent-based approach with a dynamic update of opinions and connections between agents and reflects opinion propagation and structural changes over time as observed in real-world data. We validate the model using data from the Social Evolution dataset of the MIT Human Dynamics Lab describing changes in friendships and health self-perception in a targeted student population over a nine-month period. We demonstrate the effectiveness of the approach by predicting changes in both opinion spread and connectivity of the network. We also use the model to evaluate how the network parameters, such as the level of `openness' and willingness to incorporate opinions of neighboring agents, affect the outcome. The model not only provides insight into the dynamics of ever changing social networks, but also presents a tool with which one can investigate opinion propagation strategies for networks of various structures and opinion distributions.

\end{abstract}


\clearpage

\addtocounter{page}{-2}

\pagestyle{fancy}

\section{Introduction}

People in real-world societies interact with each other, and as a result, change both their opinions and social connections. Understanding and predicting these changes is crucial in applications such as predicting disease spread \cite{disease}, understanding rumor propagation through the societies \cite{rumor}, and organizing marketing campaigns \cite{marketing}. Social networks are often modeled as graphical structures of connected agents along with their opinions. The agents of a network interact with each other stochastically and can change both their opinions and connections, often forming clusters of more tightly connected agents.

The literature dedicated to modeling social networks well represents various dynamics used to simulate agent interactions \cite{MainDiffusion, OpinionDynamics, Bradwick,  bcf, empirical, slowlearningmodel}. However, much of the literature is dedicated to models that either focus solely on the update of opinions \cite{MainDiffusion, Bradwick, bcf, slowlearningmodel} or on the update of connections \cite{pnas,  stat, empirical, upenn}. In real-world networks it is often the case that both opinions and connections of agents change over time. To reflect such complexity, we propose a social network modeling approach that incorporates changes and observed dynamics of both opinions and connections, and is based on assumptions developed through real-world data observations. Using data from the Social Evolution Dataset of the MIT Human Dynamics lab \cite{Data}, we show that the social network of students based on both self-reported friendships and opinions on a number of subjects such as health self perception exhibits both opinion and connection change over time. Additionally, we observe that the network creates clusters of more interconnected people whose opinions become more homogenized with time.

Some of the existing work using The Social Evolution Dataset \cite{Data} concentrated on tracking the evolution of social relationships using frequency of interactions \cite{DataEx1}, focusing on the importance of face-to-face interactions in causing opinion change \cite{DataEx2}, and analyzing the correlation between the duration of exposure to certain opinions and the adaptation of those opinions \cite{DataEx3}. We extend this analysis by incorporating findings in a more general model of  interactions between subjects, namely, reflecting the overall trends in opinion and connectivity over time as observed in the data. We analyze the effect of cluster formation on opinion propagation. The creation of such clusters, caused by temporal updates in connectivity of the network, has been shown to naturally form in societies \cite{pnas,  stat, upenn}, and has been observed in the data.  Several studies have accounted for clustering in social networks as a result of varied strength of social ties \cite{OpinionDynamics, upenn}. We follow this trend by using clusters as a measure of the strengths of connections between agents. 

The model provides insight into the dynamics of ever changing social networks, as well as presents a tool with which one can investigate opinion propagation strategies for networks of various structures and opinion distributions.

\section{Social Evolution Dataset} 
\label{2}
\subsection{Data Description}
\label{2.1}
We use the Social Evolution Dataset of the MIT Human Dynamics Lab \cite{Data}, which details a study of students living in a Harvard dormitory. The students\footnote{One student, (28 in the data) was a significant outlier and thus not considered in the study.} were surveyed five times over the course of eight months and asked to self-report their perceptions of various aspects of their lives such as their health, their weekly hours of exercise, political opinion, and number of fruits and vegetables consumed weekly as well as to indicate their close friends and socializing partners in the dormitory. The number of students varied from 65 in the first survey to 60 in the last. There was no demographic data such as age and gender of the students accompanying the surveys. We use data from five survey times: September 2008, October 2008, December 2008, March 2009, April 2009, denoted as: 2008.09, 2008.10, 2008.12, 2009.03, and 2009.04 respectively. In further discussion, we use the part of the survey covering each student's opinion of his or her own health (referred to as ``health opinion") and self-reported list of close friends.

\begin{figure}[h] 
\centering
\captionsetup{font=scriptsize}
\includegraphics[scale=.28]{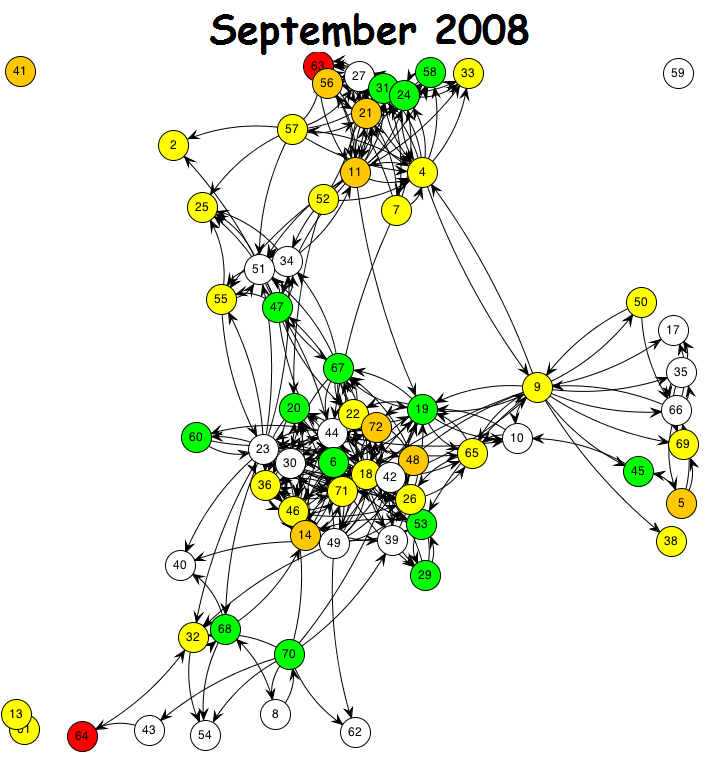}
\includegraphics[scale= .28]{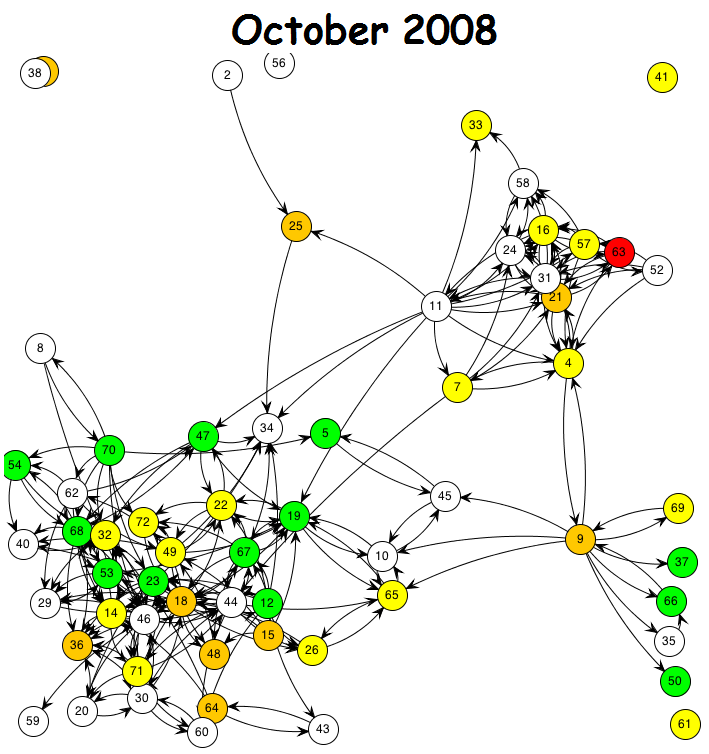}
\includegraphics[scale=.28]{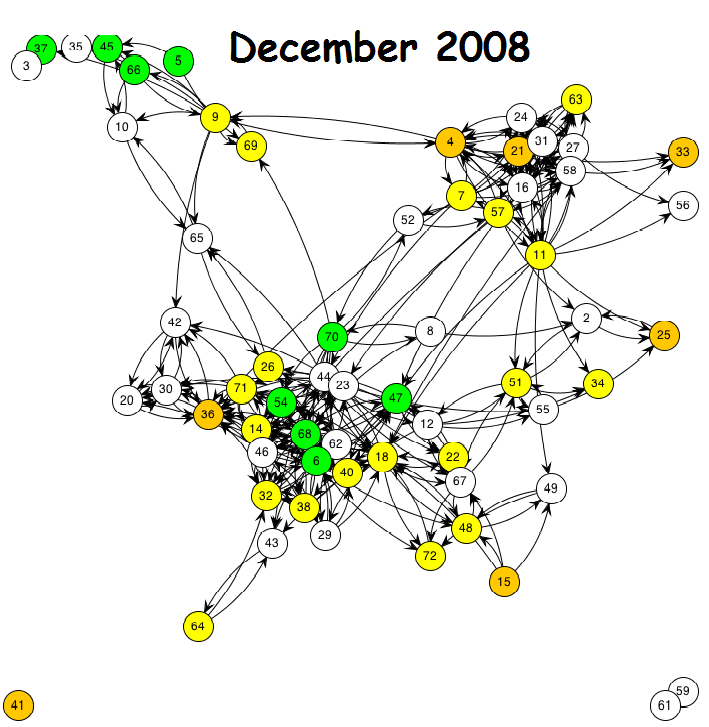}
\includegraphics[scale=.28]{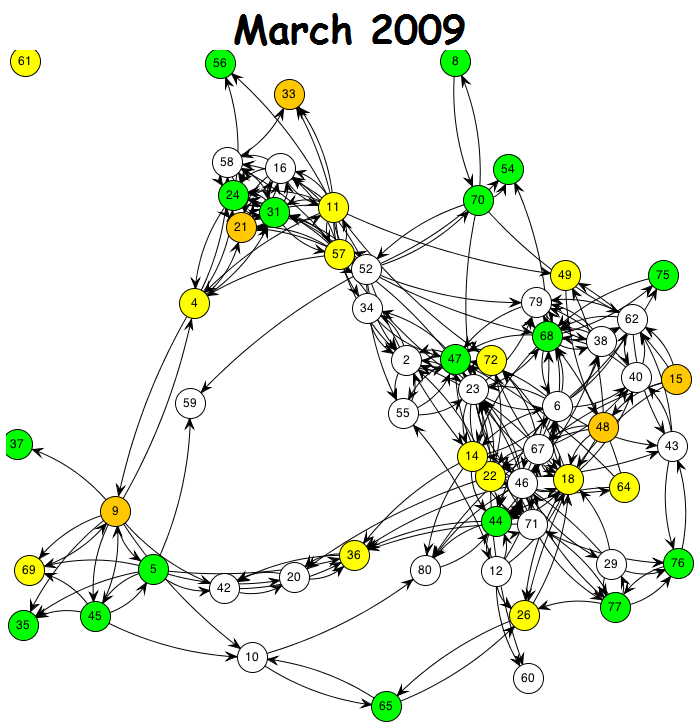}
\includegraphics[scale=.28]{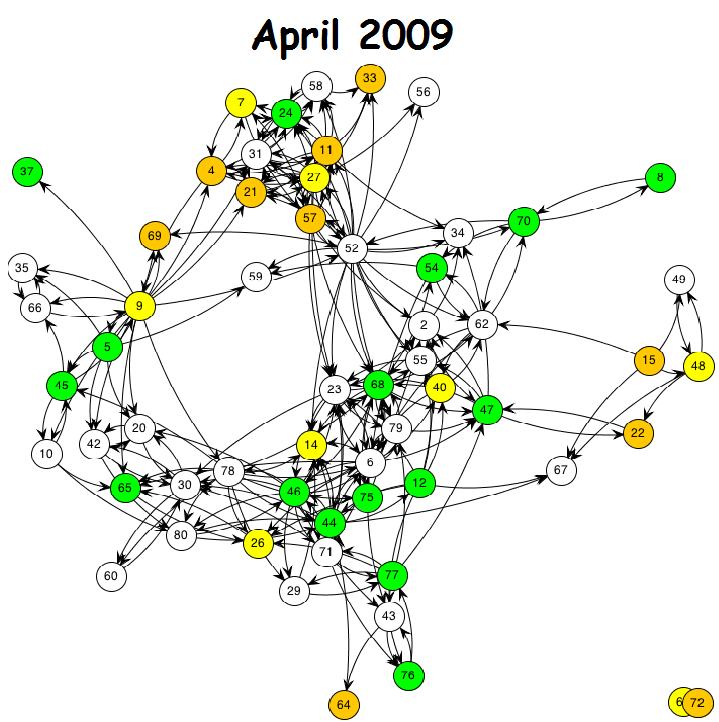}
\caption {The network representation of survey data. Each student is represented as a node in the graph; friendships are shown by directed edges. Colors reflect node opinion: healthy (green), average (white), below average (yellow), unhealthy (orange), very unhealthy (red). Note the prominent changes in opinion distribution and connectivity of the network over time.}\label{5networks}
\end{figure}
 
Health opinion as recorded in the survey could take one of five values: 2 (healthy), 1 (average), 0 (below average), $-1$ (unhealthy), and $-2$ (very unhealthy) and could be changed upon the completion of each new survey. For modeling purposes we will use a value in the continuous range $[-2, 2]$ reflecting health opinion.  \autoref{5networks} shows the change in both health opinion and friendships over time from the survey data, where each node represents a student\footnote{The graphs shown were produced using a force directed layout as documented by the JUNG library \cite{JUNG}.}. The connections in the graphs correspond to the friendships between students as indicated in the surveys. The edges are directed because in the data, friendship is not necessarily reciprocal (if student A considers student B a friend, student B does not necessarily consider student A a friend). The colors in the graphs correspond to the self-perceived health values reported by the students. As reflected in the graphs, students change both their opinions and friendships between surveys, as well as form distinct clusters. The tendency for agents in social networks to naturally form clusters was observed in several studies  \cite{pnas, upenn, stat}.

Below we discuss an approach to identify clusters in the network. There we use an undirected version of the graph\footnote{If both edges $(i, j)$ and $(j, i)$ exist in the directed graph, then $i$ and $j$ are connected by an undirected edge. Otherwise, no edge exists between $i$ and $j$ in the undirected version.} in order to increase the performance of the clustering algorithm and produce more dense clusters. 

\subsection{Cluster and Data Evaluation Metric}
\label{2.2}
We analyzed dynamic trends in the average opinion and connectivity of clusters in the student network. We define clusters, highly interconnected groups of agents that share few connections with the rest of the network, using a variant of the algorithm proposed by Girvan and Newman \cite{Algorithm} as implemented by the JUNG library \cite{JUNG}. We determine the optimal set of clusters based on the following metric, $m$, proposed by Schaeffer \cite[pg. 38]{metric}: 
\begin{equation}\label{clustermetric}
m = \frac{deg_{int}}{deg_{int} + deg_{ext}},
\end{equation}
in which $deg_{int}$ is the number of edges connecting two nodes within the given cluster and $deg_{ext}$ is the number of edges connecting a node within the given cluster to a node outside the given cluster. The metric implies that in a high quality cluster\footnote{We define a cluster as a set of nodes with size greater than 2. Throughout the paper, only clusters satisfying this property are considered in our modeling and analysis}, nodes share many connections with each other and few connections with the rest of the network. Using \eqref{clustermetric}, we compute the average quality of each set of clusters produced by the clustering algorithm, and select the best subset based on the highest average value of $m$. The defined optimal set is used in further analysis. 

Let the nodes in cluster $C$ of size $k$ be $c_1, c_2, \dots , c_k$ and have opinions $o(c_1), o(c_2), \dots , o(c_k)$ respectively. For each cluster, we introduce the following measures:

{\bfseries Cluster Opinion:} We define cluster opinion, $O(C)$, as the average opinion of the nodes in the cluster: \begin{equation}\label{opinion}O(C) = \frac{\sum_{1 \le i \le k}^{}\ o(c_i)}{k}.\end{equation}

{\bfseries Opinion Spread:} We define opinion spread, $S(C)$, as a measure of how far, on average, the opinions of the nodes in the cluster are from the cluster opinion: \begin{equation}\label{spread}S(C) = \frac{{\sum_{1 \le i \le k}^{}\ \left |o(c_i) - O(C)\right|}}{k}.\end{equation}

{\bfseries Inner Connectivity:} We define the inner connectivity, $I(C)$, of a cluster as the number of edges in the cluster divided by the number of possible edges: \begin{equation}\label{connectivityeq}I(C) = \frac{\left| E \right|}{k(k - 1)}.\end{equation} Where $\left| E \right|$ is the number of directed edges in the cluster.

\subsection{Data Observations}
\label{2.3}
We cluster the network at each of the five survey times and for each survey time, compute the average value of the proposed cluster metrics. The results are summarized in \autoref{table}, reflecting trends on both opinion and connectivity change over time.

\begin{table}[h]
\centering
\captionsetup{font=scriptsize}
\begin{tabular}[t]{| >{\columncolor[gray]{.9}}c | c | c | c | c | c | c | c |}
\multicolumn{8}{c}{}\\ \hline \rowcolor[gray]{.9}
 & 2008.09& 2008.10 & 2008.12 & 2009.03 & 2009.04 & trend & p-val\\ \hline
Cluster Opinion & .57 & .67 & .68 & 1.03 & 1.02 & \color{ForestGreen} increases & \color{cyan} .00402\\ \hline
Opinion Spread & .87 & .77 & .66 & .59 & .52 & \color{red} decreases & \color{cyan} .00176\\ \hline
Inner Connectivity & .48 & .60 & .63 & .65 & .66 & \color{ForestGreen} increases & \color{cyan}.03986\\ \hline
Avg. Cluster Size & 8.20 & 7.12 & 5.67 & 4.60 & 4.67 & \color{red} decreases & \color{cyan}.00456\\ \hline
\end{tabular} \\
\caption{The observed changes in clusters for each survey time. The values in the table are the averages of each metric over all clusters for each time set.}\label{table}
\end{table}

Trend observations were validated with a one-tailed linear regression $t$-test with time as the independent variable and the cluster metric as the dependent variable. As the average cluster size decreases over time, the connectivity within cluster increases, and opinions become more uniform. We observed an overall increase in the value of health self-perception over time:

{\bfseries Observation 2.1:} The average cluster opinion significantly increases over time $(p = .00402)$.

{\bfseries Observation 2.2:} The average opinion spread of the clusters significantly decreases over time $(p = .00176)$.

{\bfseries Observation 2.3:} The inner connectivity significantly increases over time $(p = .03986)$.

In addition to Observation 2.1, analysis of individual opinions showed a pronounced tendency towards moving away from both the extreme negative opinion (-2) and the neutral one (0) as seen in \autoref{histograms}. In 2008.09 there were 22 students with opinion 0 while in 2009.04 there were only 8 such students. Furthermore, by 2009.04, 42 students adopted positive opinions while only 10 students adopted a negative opinion.  The movement away from negative opinions towards positive ones is further supported by the fact that in 2009.04, there are 0 students with opinion $-2$, the most extreme negative opinion, while there are 16 students with opinion 2, the most extreme positive opinion. The model proposed in the next sections takes these observations into account (see \autoref{3}):  

\begin{figure}[h]
\captionsetup{font=scriptsize}
\includegraphics[scale=.45]{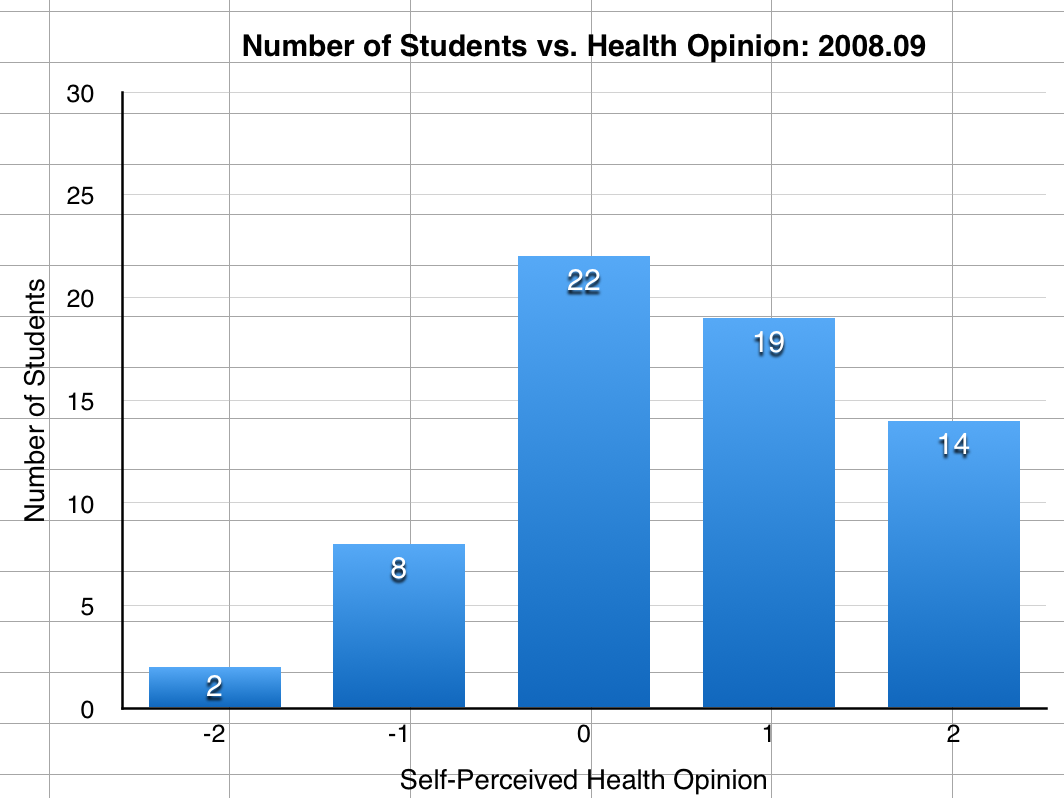}
\includegraphics[scale=.45]{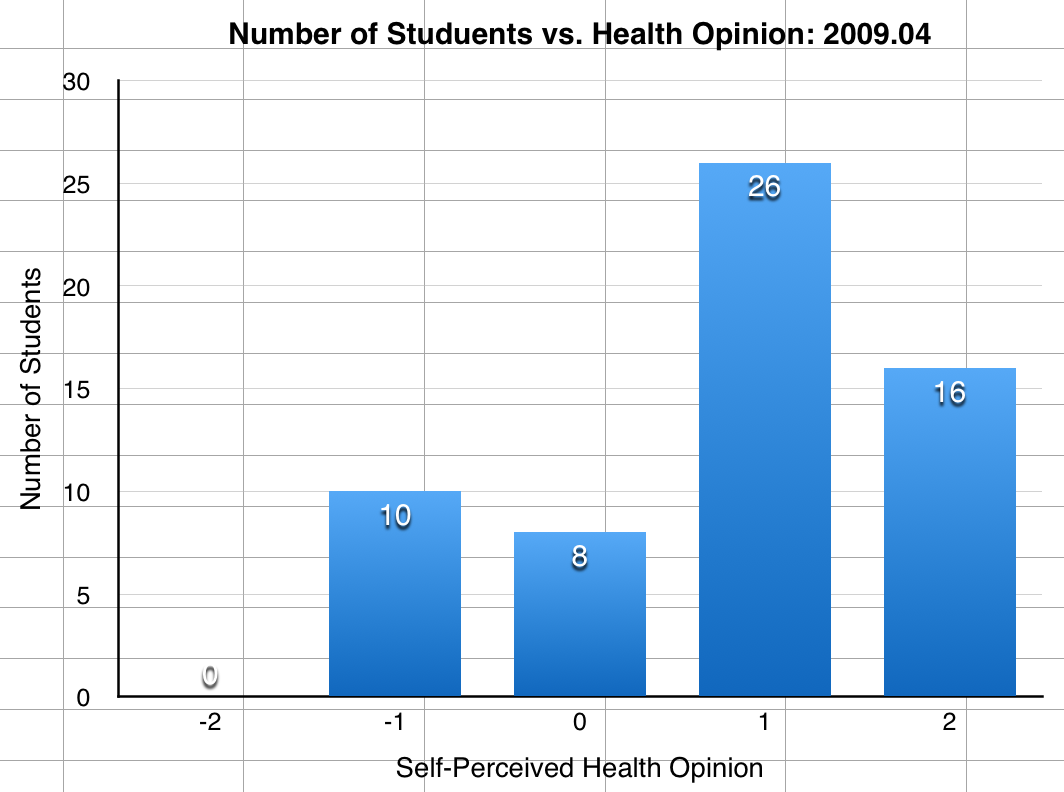}
\caption{Histogram of self-perceived students' health for survey times 2008.09 (left) and 2009.04 (right). Note the significant deviation away from 0, the neutral opinion value, towards more opinionated values such as -1, 1, and 2. The number of agents that reach strong positive opinions (opinions of 1 and 2) is significantly higher than the number of agents that reach strong negative opinions (opinions of -1 and 2), suggesting  an asymmetric bound}\label{histograms}
\end{figure}

{\bfseries Observation 2.4:} Individual opinions tend to move away from the neutral opinion value. The observed shift in positive direction is more significant than that in the negative direction.

Based on the above-mentioned observations we conclude that in addition to overall tendency to move opinion in a more positive directions, students have a tendency to `amplify' their opinion. Particular enhancement of opinion is observed when it falls within $(-1, 0) \cup (1.5, 2)$. This range of values and observed asymmetry will be used in model simulations (see \autoref{4}).

\section{Agent-Based Model of the Dynamic Network}
\label{3}
Using observations from data, we develop an agent-based model \cite{agentbased, upenn} aimed at both recreating and predicting the trends in both cluster opinion and connectivity observed in Social Evolution Dataset (\autoref{2}). The model implements the changing of both opinions and connections via a stochastic process, in which the state of the network at time $t + 1$ is determined by the state of the network at time $t$. 
\subsection{Structure of the Network}

Define the network at time $t$ as a directed graph $G(N, A_t)$ with set of nodes $N$ and adjacency matrix $A_t$, denoting friendships between agents. Each node $i$ in the network is an agent having an opinion $o_i(t)$ at time $t$. 

As observed in the data, $A_t$ can be asymmetric. In addition, several studies have demonstrated that social ties often have unequal strength \cite{OpinionDynamics, upenn}. We define the difference in strengths of connections between nodes through cluster memberships, so that connections between nodes within a cluster are stronger than those between nodes from different clusters. Using this concept, we define $A_t$ as follows: 
 \begin{equation}\label{weights}a_{ij}(t) =   
\begin{cases}
  0 &  \text {if no edge connects node } i \text{ to node } j \text{ (no $i$ to $j$ friendship),} \\
  1 &  \text  {if  } i \text{ is connected to } j \text{ from a different cluster ($i$ befriends $j$),}\\
  w \text{ } (w \ge 1) & \text {if  } i \text{ is connected to } j \text{ from the same cluster ($i$ befriends $j$),}
 \end{cases} \end{equation}
where $a_{ij}(t)$ denotes row $i$, column $j$ of $A_t$. The parameter $w$ reflects the strength of cluster effect (i.e. how much stronger friendships are between agents in the same cluster versus between agents in different clusters).

\subsection{Opinion Update}
\label{3.2}

At each time $t$, agents interact with each other through existing friendship connections. As a result, each node's opinion is updated according to the influence exerted by its friends. In particular, we assume that interaction with friends affects opinion change, and that a friend's influence is stronger if both agents belong to the same cluster. We also take into account the opinion amplification effect discussed in Section 2. 

Let $q_i(t)$ be the influence that agent $i$'s friends exert on agent $i$'s opinion, $o_i(t)$: \begin{equation}\label{q(x)}q_i(t)= \sum_{j \in N}^{}(\alpha) \frac{a_{ij}(t)}{\sum_{k \in N}^{} a_{ik}(t)} (o_j(t) - o_i(t)),\end{equation} where $\alpha \in (0, 1]$ is a slow learning scale factor \cite{slowlearningmodel}. The equation reflects the assumption that all the friends affect the agent's opinion, but that the effect is proportional to the strength of connection between agents.  

In addition, we take into account the tendency of agents to amplify their opinions towards a more extreme value. As noted in Observation 2.4 and supported by \cite{radicalization2, radicalization1}, opinions in social networks tend to radicalize toward more extreme values over time. In other words, the longer an agent possesses any given opinion, the stronger that opinion becomes. We introduce an opinion amplification function, $s(y)$ to reflect this observation:
\begin{equation}\label{amp}s(y) =   
\begin{cases}
  ky \text{ } (k \ge 1)  & \text{if }y\in D,\\
  y &  \text {otherwise,}
\end{cases}\end{equation}
where $D$ is an empirically defined domain of opinions in which such amplification takes place and $k$ is a parameter representing the extent to which agents have a natural tendency to amplify their opinions over time. 

Taking \eqref{q(x)} and \eqref{amp} into account, we define the opinion $o_i(t+1)$ of every agent $i$ as follows: \begin{equation}\label{updateRule}o_i(t + 1) = s(o_i(t) + q_i(t)).\end{equation}

\subsection{Connection Update}

The update of connections follows the update of opinions from time $t$ to time $t + 1$. After the update of opinions, each node will probabilistically form one connection and probabilistically break one connection using the following assumptions: 

{\bfseries 3.3.1:} Agent $i$ can form friendships with friends of its friends \cite{stat, upenn} 

{\bfseries 3.3.2:} Agent $i$ can form friendships with agents that already consider him a friend \cite{pnas, stat, ncbi}.

{\bfseries 3.3.3:} The stronger the potential connection between two disconnected  agents, the more likely the connection is to form \cite{OpinionDynamics, upenn}. In terms of our model, a connection between agents within the same cluster is more likely to form.

{\bfseries 3.3.4:} The stronger the connection between two connected agents, the less likely the connection is to break \cite{breaking}. In terms of our model, a connection between agents in the same cluster is less likely to break. 
 
We define $S_f$ as a set of edges that could form, and  $S_b$ as a set of edges that could break. By definition, $S_f \cap S_b = \emptyset$.

Define probability $f_{ij}(t)$, the probability that a directed connection from $i$ to $j$ forms between times $t$ and $t + 1$, and $b_{ij}(t)$, the probability that a directed connection from $i$ to $j$ breaks between times $t$ and $t + 1$ as follows: \begin{subequations}
\begin{align}\label{connection_probabilities_a}f_{uv}(t) =
\begin{cases}
  \beta p_{ij}(t) &  \text {if  }(i, j) \in S_f,\\
  0 & \text{if }(i, j)\not\in S_f,
\end{cases}  \end{align} \begin{align}\label{connection_probabilities_b}
b_{ij}(t)=   
\begin{cases}
  \beta(1 - p_{ij}(t)) &  \text {if  }(i, j) \in S_b,\\
  0 & \text{if }(i, j)\not\in S_b,
\end{cases}
\end{align}
 \end{subequations}respectively, where $\beta \in (0, 1]$, and $p_{ij}(t)$ is:
\begin{equation}\label{p(x, y)}p_{ij}(t) = 
\begin{cases}
  (.5 - c) &  \text {if } i \text{ and } j \text{ are not in the same cluster,}\\
(.5 + c) &  \text {if } i \text{ and } j \text{ are in the same cluster,}\\ 
 (.5) &  \text {if } j \text{ is in not in any cluster} 
 \end{cases} \end{equation} The parameter $c \in [0, .5)$ represents the extent to which agents restrict their friendships to those within their own cluster. Function $p_{ij}(t)$ accounts for the assumption that connections are more likely to form and stay within clusters rather than between clusters. 
Using \eqref{connection_probabilities_a}, \eqref{connection_probabilities_b}, and \eqref{p(x, y)}, we update the adjacency matrix from $A_t$ to $A_{t + 1}$ using the Monte Carlo method \cite{MonteCarlo}.  The network is then re-clustered using the algorithm in \autoref{2} and $A_{t + 1}$ is updated again according to \eqref{weights}.

\section{Results}
\label{4}

\subsection{Modeling and Predicting Network States in the Social Evolution Dataset}
\label{4.1}

In this section, we apply the proposed model to the Social Evolution Dataset in order to demonstrate the model's effectiveness in replicating the observed trends of both opinion spread and connectivity. We use the earlier defined measures of cluster opinion, opinion spread, and inner connectivity in order to compare model results with the observed data.  We recreate the short term trends observed in the data, and extend projections to future states thus anticipating longer-term network dynamics. 

\begin{figure}[h]
\includegraphics[scale=.32]{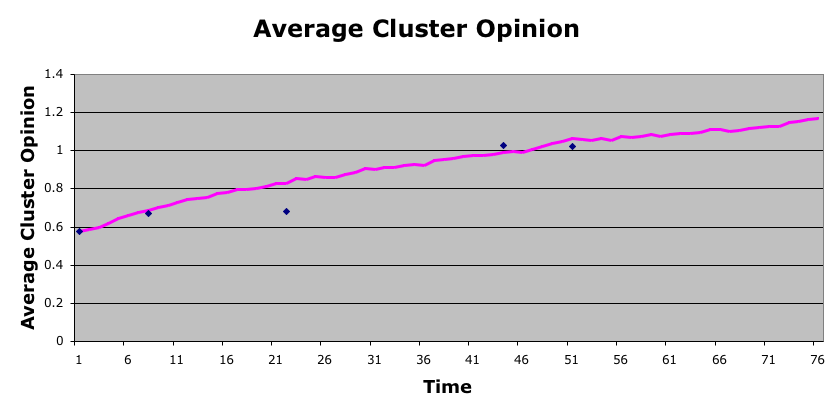}
\includegraphics[scale=.32]{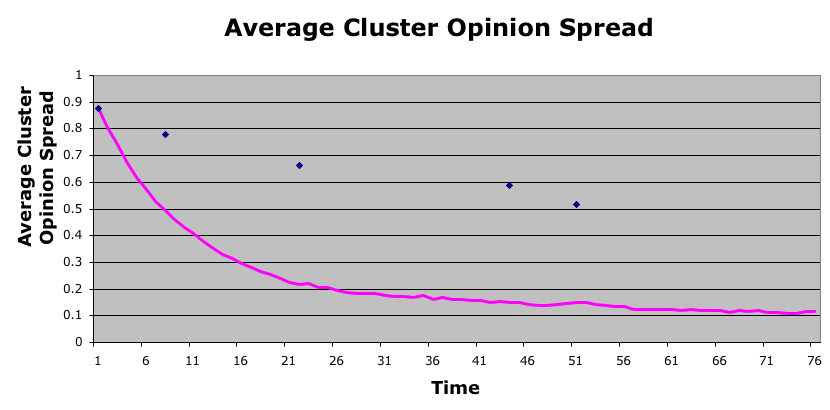}
\begin{minipage}[c]{0.30\textwidth}
\includegraphics[scale=.32]{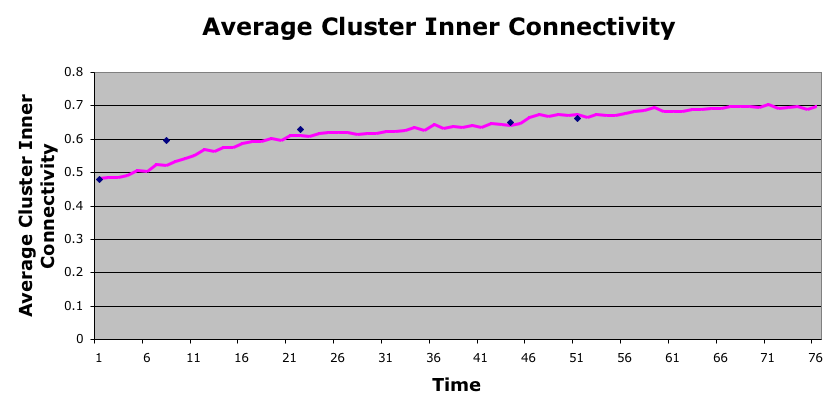}
  \end{minipage}\hfill
  \begin{minipage}[c]{0.3\textwidth}
   \captionsetup{font=scriptsize}
   \caption{Graphs of the model output (pink) and the data (blue). The model curves follow the trends in the data points for the three metrics defined in \eqref{opinion}, \eqref{spread}, and \eqref{connectivityeq}.}\label{datavalidationshort}
  \end{minipage}
\end{figure}

In application to the Social Evolution Dataset we define $D$ from \eqref{amp} as  $(-1, 0) \cup (1.5, 2)$ (see Observation 2.4). We also bound opinion values produced by \eqref{amp}  to $[-2, 2]$ in accordance with data (\autoref{2}), setting any value less than $-2$ or greater than 2 to $-2$ and 2 respectfully. 

The model has three parameters: $w$ (tendency to follow the opinion of those from the same cluster versus those from a different cluster), $k$ (tendency to radicalize opinion), and $c$ (tendency to form more friendships with those from the same cluster versus with those from a different cluster), along with two scalars, $\alpha$ and $\beta$, which regulate the rate of opinion and connection change respectively. Using the network from the first survey time in our data (2008.09) as an initial condition ($t = 0$), we ran simulations in order to identify values for the parameters that best describe the social network dynamics of the data.

\autoref{datavalidationshort} shows the results of simulations using parameter values $w = 5$, $k = 1.05$, $c = .245$ and scalar values $\alpha = .10$, $\beta = .15$ (pink curve), which were selected for providing the best fit to the data (shown as the blue points). The results shown are the average over 50 simulations. 

The accuracy of the model was evaluated using the percent error\footnote{percent error = $\frac{\left| \text{model value} - \text{data value}\right|}{\text{data value}} * 100\%$}. Cluster opinion had an average percent error of $7.07\%$ and inner connectivity had an average percent error of $5.22\%$, thus indicating that the model is accurate within a reasonable error bound in predicting the trends in these two metrics. The opinion spread generally followed the observed trend, though the percent error was significantly larger. Note that while the model's prediction of cluster opinion closely followed the data trend, the opinion spread decreased more quickly in the model than in the data. This means that the average opinion of the clusters in the model changed at the same rate as the average opinion of the clusters in the data but the opinions of the agents in each cluster converged to the average opinion of the cluster faster in the model than in the data.

\begin{figure}[h]

\captionsetup{font=scriptsize}
\includegraphics[scale=.50]{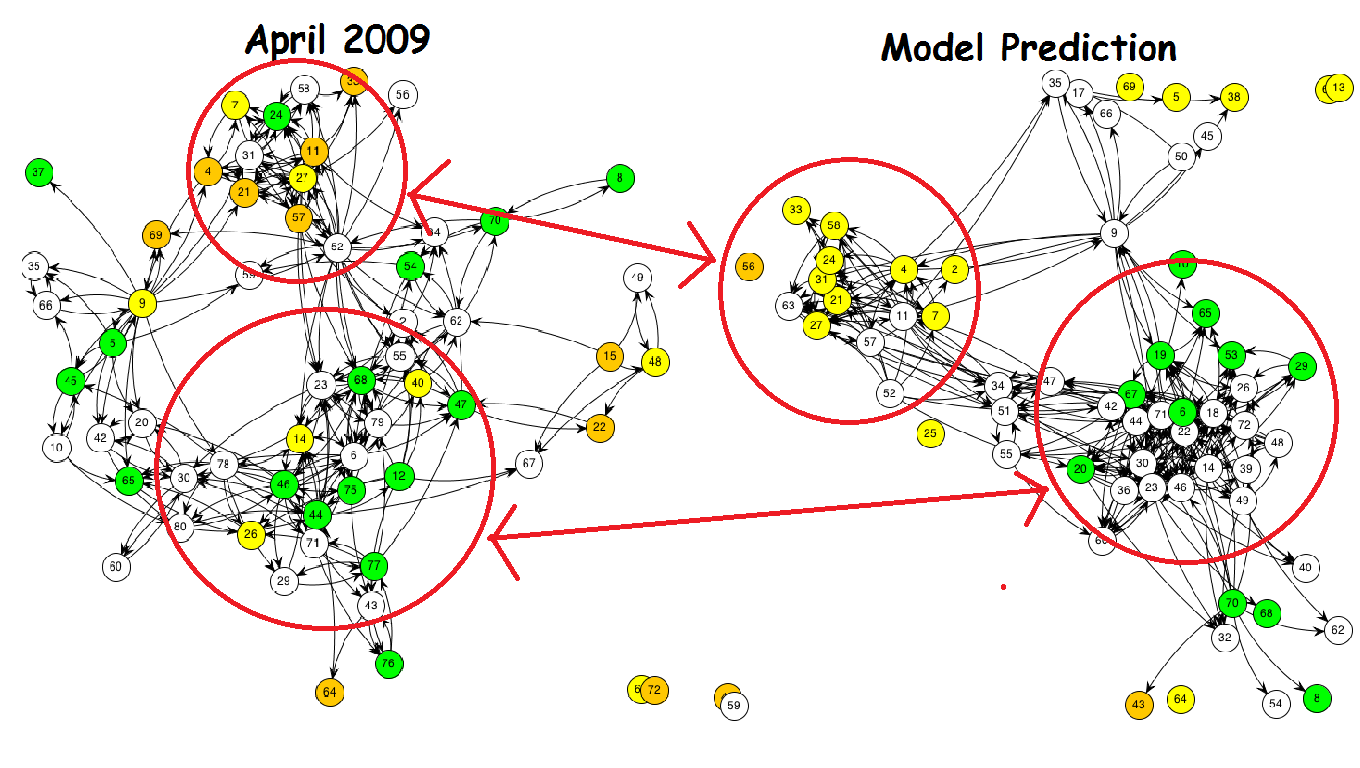}
\caption{The data at 2009.04 (left) and the model at $t = 50$ (right). Note the similarity between the two networks in the existence of a larger cluster that is mostly green (opinion value 2) and white (opinion value 1) and a smaller cluster that is mostly yellow (opinion value 0) and orange (opinion value $-1$). Visuals created with assistance of code from the JUNG library. \cite{JUNG} \label{datanetworks}}
\end{figure}

At t=50 in the simulations, the average opinion in the model is close to 1.0 and the average inner connectivity is close to .65 (\autoref{datavalidationshort}). These values are close to those observed in 2009.04 survey, the last survey in the data (see \autoref{table}). \autoref{datanetworks} further demonstrates the similarity between the network output at $t = 50$ and the data in 2009.04. Note the similarity of the two networks in the existence of two prominent clusters, one containing agents with positive opinions (white and green) and one containing agents with neutral and negative opinions (yellow and orange). 

\begin{figure}[h]
\includegraphics[scale=.32]{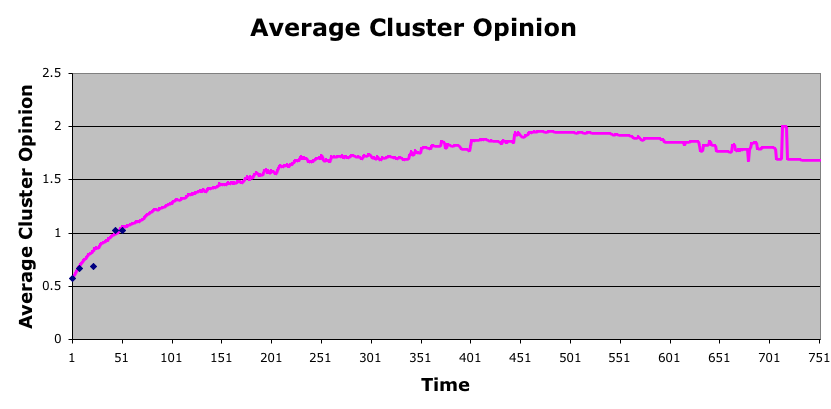}
\includegraphics[scale=.32]{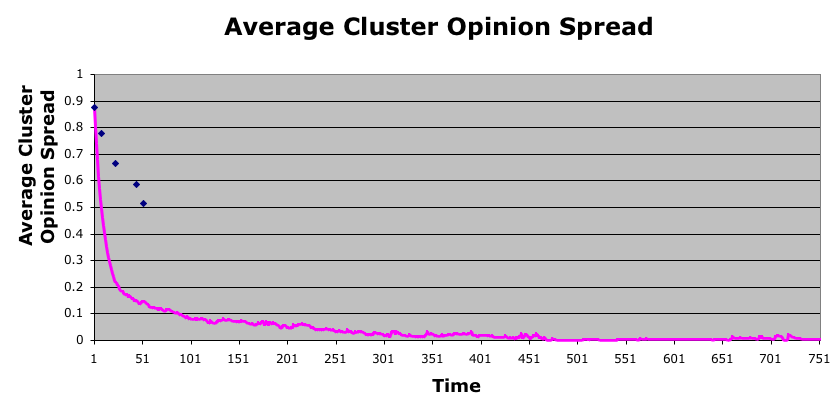}
\begin{minipage}[c]{0.30\textwidth}
\includegraphics[scale=.32]{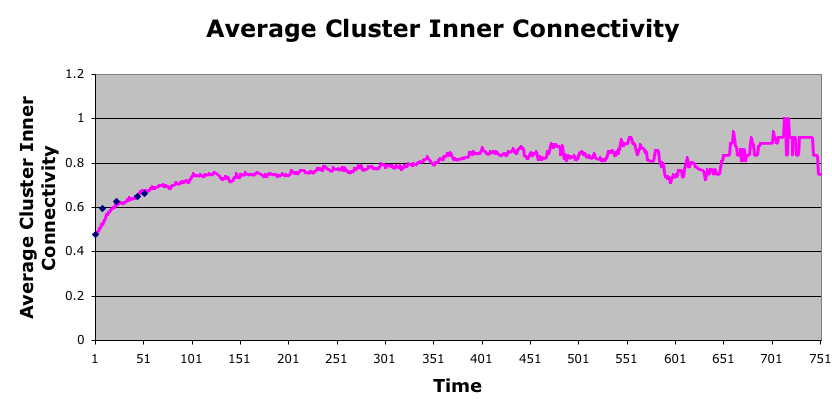}
  \end{minipage}\hfill
  \begin{minipage}[c]{0.3\textwidth}
   \captionsetup{font=scriptsize}
   \caption{Graphs of the model output (pink) and the data (blue). The model curves follow the trends in the data points for the three metrics defined in \eqref{opinion}, \eqref{spread}, and \eqref{connectivityeq}. The graphs both reflect the short term trends we observed in the data and predict the long term changes in the network.}\label{datavalidationlong}
  \end{minipage}
\end{figure}

\autoref{datavalidationlong} shows longer term simulation results using the same initial conditions and set of parameters. As time increases, the average cluster opinion spread approaches 0, which suggests that with this set of parameters, the opinions of nodes in individual clusters will converge to the same value, between 1.5 and 2 (\autoref{datavalidationlong}). 

\subsection{Network Dynamics Under Varying Social Conditions}
\label{4.2}

The parameters in the model represent a number of aspects of society such as tendency to follow the opinion of those from the same group (cluster) versus those from a different group (parameter $w$), tendency to amplify (`radicalize') opinion (parameter $k$), and tendency to form new friendships with those from different groups rather than the same group (parameter $c$). We vary the parameters in order to gain an understanding of how the network dynamics change under different sets of conditions governing interactions between agents.\footnote{Note however, that we do not consider the constants $\alpha$ and $\beta$ parameters. Instead, they are data-driven constants used to control the speed at which opinion and connection update respectively occur for which the values $\alpha = .10$ and $\beta = .15$ were confirmed in \autoref{4.1}.}

We use the first survey data (2008.09) as the initial condition for the model at time $t = 0$ and vary only one parameter at a time in order to avoid the confounding effects. Note that varying $w$ and $k$ only affects changes in cluster opinion and opinion spread as these two parameters have no effect on the changing of connections. When varying $c$, changes are observed in the inner connectivity of clusters, followed by changes in cluster opinion and cluster opinion spread. 

\subsubsection{Clustering effect on opinion change}
\begin{figure}[h]
\captionsetup{font=scriptsize}
\includegraphics[scale=.32]{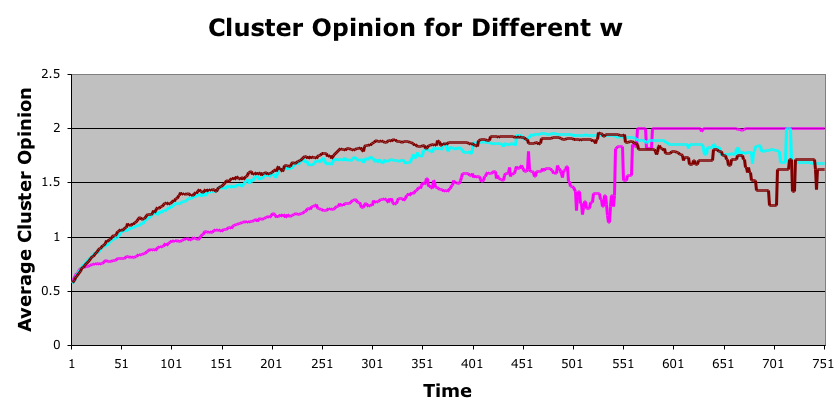}
\includegraphics[scale= .32]{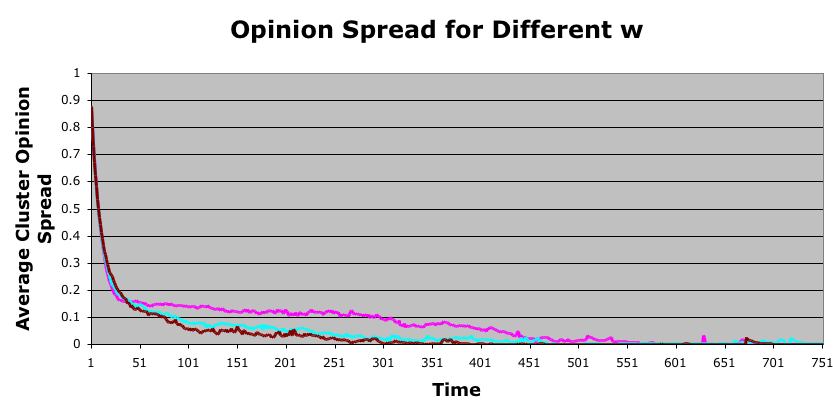}
\caption{Average cluster opinion (left) and average cluster opinion spread (right) for $w = 1$ (pink), $w = 5$ (blue), and $w = 100$ (brown). We hold $c = .245$ and $k = 1.05$ constant. Note that as $w$ increases, the differences between the three curves in long term average cluster opinion becomes negligible, thus suggesting that weighted friendships only affect long term opinion propagation up to a certain point.}\label{w}
\end{figure}

Giving preference to agents in the same cluster does not affect the long-term average opinion and opinion spread past a certain degree of preference. However, a low degree of preference significantly affects a rate of convergence to long-term cluster opinion (see \autoref{w}). 
 
{\bfseries 4.2.1.1:} For large values of $w$, there is no observable difference in both long-term opinion dynamics and opinion spread dynamics (\autoref{w}). 

High values of $w$ indicate that agents are influenced to a greater degree by agents within their cluster than by those outside one. We ran 50 simulations each for 5 different values of $w>1$ between 5 and 100 (with $w = 5$ and $w = 100$ shown in \autoref{w}) and observed a convergence within $12.5\%$  after 50 iterations. We conclude that given that agents follow the opinions of friends within their own cluster ($w > 1$) more, the degree to which they do so does not significantly impact cluster opinion and cluster opinion spread in the long run. 

{\bfseries 4.2.1.2:} A slower rate of change occurs in both average cluster opinion and opinion spread when agents have no tendency to favor the opinions of their cluster members over the opinions of the rest of the network ($w = 1$).

 When $w = 1$, all agents in the network have an equal influence on any given agent's opinion, regardless of the clusters they belong to (equations \eqref{weights} and \eqref{q(x)}). As discussed in Section 2, the average cluster opinion of the network is initially positive, with more agents having positive opinion than negative one. In combination with amplification effect (see \eqref{amp}) this leads to a slow increase in the average opinion value over the network, surpassing the effect of negative amplification. 

\subsubsection{The tendency to amplify (`radicalize') opinions}

When varying opinion amplification parameter $k$, we analyzed the effect of amplification of opinion (`radicalization') on overall network opinion and opinion spread.  

\begin{figure}[h]
\captionsetup{font=scriptsize}
\includegraphics[scale=.32]{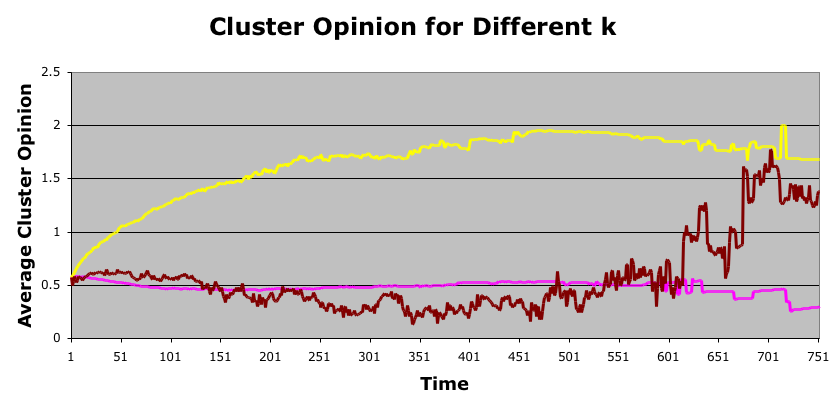}
\includegraphics[scale=.32]{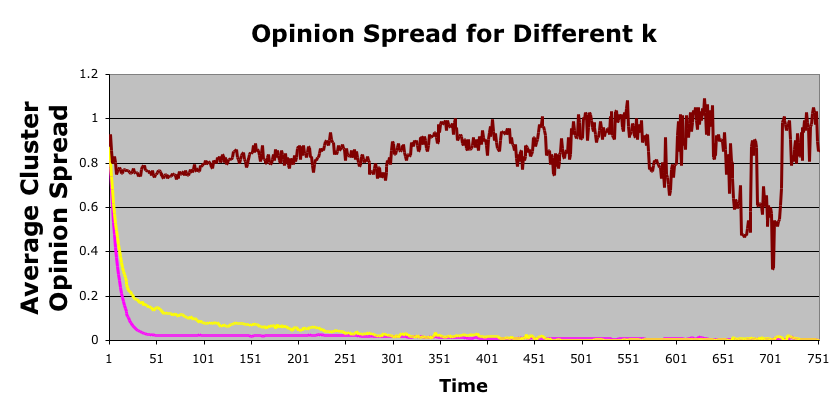}
\caption{Average cluster opinion (left) and average cluster opinion spread (right) for $k = 1$ (pink), $k = 1.05$ (yellow), and $k = 2.00$ (brown) while holding $c = .245$ and $w = 5$ constant. Note the long term instability of the curves for $k = 2.00$ which results from very large opinion amplification.}\label{k}
\end{figure}

{\bfseries 4.2.2.1:} Average network opinion remains fairly constant and opinion spread decreases most rapidly when agents have no tendency to amplify opinions ($k = 1.00$). 

With no opinion radicalization, agents' opinions are not pulled away from the average opinion of the network towards more extreme opinion values. Therefore, the average cluster opinion remains fairly constant and the opinions of agents in the network converge towards the average very quickly.

{\bfseries 4.2.2.2:} Average network opinion increases significantly and opinion spread decreases more slowly when agents have a slight tendency to radicalize opinions ($k = 1.05)$. 

When $k = 1.05$, there is a slight radicalization of opinions. From the bounds we implement on the amplification function in \eqref{amp}, we know that positive opinions are favored over negative opinions. Therefore, the amplification of negative opinions is not enough to offset the amplification of positive opinions, resulting in an increase in average opinion. However, the fact that there is a slight amplification in the negative direction as well as in the positive one causes the opinion spread to decrease more slowly as the opinions in the network are pushed away from the average.

{\bfseries 4.2.2.3:} Average opinion and opinion spread are very unstable when agents have a very high tendency to radicalize opinions ($k = 2.00$). 

The amplification function defined in \eqref{amp} affects both positive and negative opinions for high values of $k$. As a result of probabilistic interactions with their neighbors, the agents' opinions are randomly pulled in both positive and negative directions. As opinions become increasingly amplified with time, the random swings become stronger, resulting in significant fluctuations of opinion seen in \autoref{k}.

\subsubsection{Increase of connectivity within clusters}

\begin{figure}[h]
  \begin{minipage}[c]{0.30\textwidth}
\includegraphics[scale=.32]{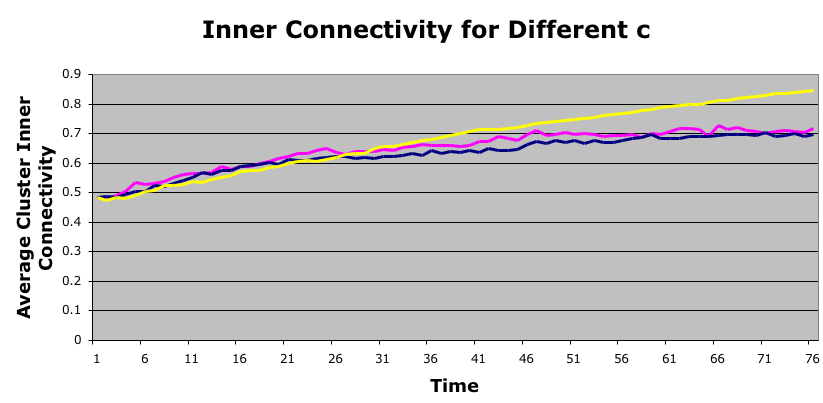}
  \end{minipage}\hfill
  \begin{minipage}[c]{0.3\textwidth}
   \captionsetup{font=scriptsize}
  \caption{Average cluster inner connectivity for $c = 0$ (pink), $c = .245$ (blue), and $c = .49$ (yellow) with  $w = 5$ and $k = 1.05$ held constant. Note that the curves for $c = 0$ and $c = .245$ level off long term while the curve for $c = .49$ tends to increase rapidly.} \label{connectivity} 
\end{minipage}\end{figure}

Parameter $c$ reflects the tendency of agents to connect with their own cluster over the rest of the network. We ran simulations with five different values of $c$. \autoref{connectivity} shows the change in inner connectivity over time for three of the values, $c = 0$, $c=.245$, and $c=.49$. There is no observed change in connectivity levels within clusters for $c=0$ and $c=.245$, while inner connectivity significantly increases for $c = .49$. This leads to a different dynamics of opinion change.

\begin{figure}[h]
\captionsetup{font=scriptsize}
\includegraphics[scale=.32]{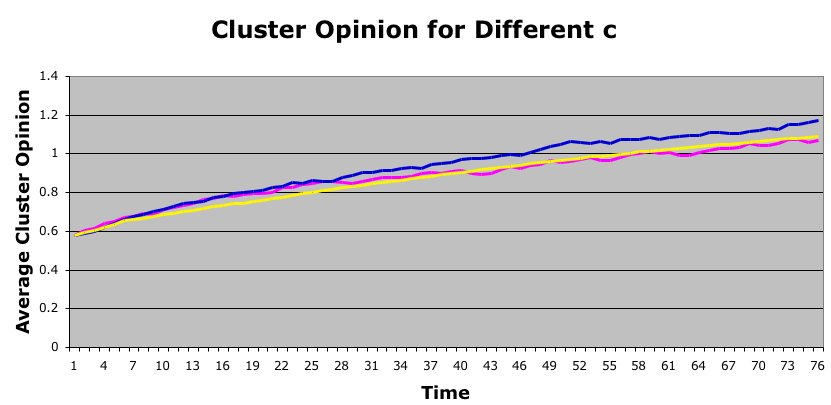}
\includegraphics[scale=.32]{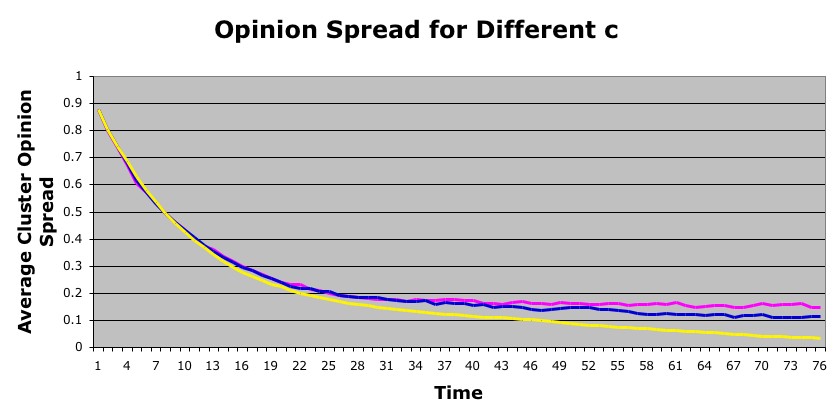}
\caption{Average cluster opinion (left) and average cluster opinion spread (right) for $c = 0$ (pink), $c = .245$ (blue), and $c = .49$ (yellow) while holding $w = 5$ and $k = 1.05$ constant. Note that as $c$ increases, average cluster opinion spread decreases more quickly while no clear trend is observed in average cluster opinion over a long period of time. The difference observed in cluster opinion spread but not in cluster opinion suggests that agents in a more interconnected cluster tend to the same opinion more quickly, but what opinion the agents tend towards is not affected by the interconnectedness of the cluster.}\label{varyc}
\end{figure}

{\bfseries 4.2.3.2:} As agents' tendency to connect with their own cluster increases, average cluster opinion spread decreases more quickly but cluster opinion is not significantly affected. 

By equation \eqref{q(x)}, a more interconnected cluster means that a given agent will be influenced by the opinions of a greater number of agents within its cluster, or in other words, there will be more interactions between agents of a given cluster. As a result of the greater number of interactions, the opinions of agents within a cluster will converge towards a similar value more quickly (\autoref{varyc}). Although $c$ influences the number of opinion updates between agents, it does not influence the nature of those interactions. Therefore, the network will tend towards the approximately the same average cluster opinion for different values of $c$, but at different rate.

\section{Conclusion}
\label{conclusion}

We have developed an agent-based social interactions model incorporating temporal changes in both opinion and connectivity through local interactions. The model successfully replicated observations from a real-world MIT Human Dynamics Lab dataset \cite{Data} containing surveys of a student population. We have identified parameter values with which we were able to reproduce general trends observed in both opinion and connectivity in data from a student network. In addition, we used the model to forecast the longer term dynamics of the observed network. We demonstrated that in order to accurately reflect trends from data, one needs to take into account not only opinion influence and interchange in the student population, but also changes in network structure reflected in the model as connections between nodes. 

We analyzed the effect of group membership on opinion change. When varying $w$, the extent to which weighted friendships due to clustering affects opinion change, we observed that weighted friendships between agents only affects long term opinion propagation up to a certain point, beyond which the long term opinion propagation changes negligibly. When $w = 1$, meaning that agents are equally influenced by all those they are connected to regardless of cluster membership, average cluster opinion increases more slowly than when $w > 1$ and average cluster opinion spread decreases more slowly than when $w > 1$. 

When varying $k$, the extent to which opinions in a human society `radicalize', we noted that a small increase in tendency to radicalize results in an increase in long-term average cluster opinion and a slower decrease in opinion spread, while a large increase in tendency to radicalize results in long-term instability in both cluster opinion and opinion spread. 

Lastly, we observed that when the tendency of agents to restrict their friendships to those within the same cluster increases (with higher values of $c$), the average inner connectivity of the clusters increases more quickly over time. We also found that as clusters become more interconnected, the opinions of agents in each cluster tend to converge to the same value at a faster rate.

The model created yields not only a method for reflecting the dynamics of and predicting the changes in a real human society, but also tools both for studying the changes that occur in various theoretical networks and for gaining insight into methods for manipulating the changes in these networks for societal benefit.

\section{Acknowledgements}
\label{9}

The authors would like to thank the MIT Human Dynamics Lab for providing the dataset \cite {Data} with which this research was conducted. We would also like to thank the MIT PRIMES program for providing the opportunity to perform this research.


\begin{thebibliography}{9}

\singlespace

\bibitem{MainDiffusion}
Acemoglu, Daron, Asuman Ozdaglar, and Ercan Yildiz.``Diffusion of innovations in social networks." Decision and Control and European Control Conference (CDC-ECC), 2011 50th IEEE Conference on. IEEE, 2011.

\bibitem{OpinionDynamics}
Acemoglu, Daron, Mohamed Mostagir, and Asuman Ozdaglar. ``State-dependent opinion dynamics." Acoustics, Speech and Signal Processing (ICASSP), 2014 IEEE International Conference on. IEEE, 2014.
 
\bibitem{DataEx2}
Anmol Madan, Katayoun Farrahi, Daniel Gatica-Perez, and Alex Pentland. Pervasive sensing to model political opinions in face-to-face networks. In Proceedings of the 9th International Conference on Pervasive Computing (Pervasive), pages 214-231, 2011.

\bibitem{DataEx3}
Anmol Madan, Sai T. Moturu, David Lazer, and Alex Pentland. Social sensing: obesity, unhealthy eating and exercise in face-to-face networks. In Proceedings of Wireless Health 2010 (WH), pages 104-110, 2010.

\bibitem{Bradwick}
 Bradwick, M. E. (2012). Belief propagation analysis in two-player games for peer-influence social networks (Doctoral dissertation, Massachusetts Institute of Technology). http://dspace.mit.edu/bitstream/handle/1721.1/72645/807215820.pdf?sequence=1
 
\bibitem{agentbased}
Carley, Kathleen M., et al. ``BioWar: scalable agent-based model of bioattacks." Systems, Man and Cybernetics, Part A: Systems and Humans, IEEE Transactions on 36.2 (2006): 252-265.
 
\bibitem{bcf} 
David Kempe, Jon Kleinberg, ƒva Tardos: Influential Nodes in a Diffusion Model for Social Networks. In Proceedings of ICALP 2005, Lisboa, Portugal.

\bibitem{radicalization2}
Deffuant, Guillaume, Frédéric Amblard, and Gérard Weisbuch. "Modelling group opinion shift to extreme: the smooth bounded confidence model." arXiv preprint cond-mat/0410199 (2004).

\bibitem{pnas}
D. Rand, S. Arbesman, and N.A. Christakis, ``Dynamic Social Networks Promote Cooperation in Experiments with Humans," PNAS: Proceedings of the National Academy of Sciences  108(48): 19193-19198 (November 2011); doi:10.1073/pnas.1108243108

\bibitem{radicalization1}
FrŽdŽric Amblard, Guillaume Deffuant, The role of network topology on extremism propagation with the relative agreement opinion dynamics, Physica A: Statistical Mechanics and its Applications, Volume 343, 15 November 2004, Pages 725-738, ISSN 0378-4371.

\bibitem{Algorithm}
Girvan M and Newman MEJ, ``Community structure in social and biological networks," Proc. Natl. Acad. Sci. USA 99, 8271-8276 (2002).

\bibitem{stat}
Handcock, Mark S., Adrian E. Raftery, and Jeremy M. Tantrum. ``Model-based clustering for social networks." Journal of the Royal Statistical Society: Series A (Statistics in Society) 170.2 (2007): 301-354.

\bibitem{marketing}
Hartline, Jason, Vahab Mirrokni, and Mukund Sundararajan. ``Optimal marketing strategies over social networks." Proceedings of the 17th international conference on World Wide Web. ACM, 2008.

\bibitem{empirical}
Kossinets, Gueorgi, and Duncan J. Watts. "Empirical analysis of an evolving social network." Science 311.5757 (2006): 88-90.

\bibitem{upenn}
Levine, S. S. \& Kurzban, R. (2006). Explaining clustering within and between organizations: Towards an evolutionary theory of cascading benefits.Managerial and Decision Economics, 27, 173-187.

\bibitem{Data}
Madan, Anmol, et al. ``Sensing the ``health state" of a community." IEEE Pervasive Computing 11.4 (2012): 36-45.

\bibitem{breaking}
McPherson, Miller, Lynn Smith-Lovin, and James M. Cook. ``Birds of a feather: Homophily in social networks." Annual review of sociology (2001): 415-444.

\bibitem{rumor}
Nekovee, Maziar, et al. ``Theory of rumour spreading in complex social networks." Physica A: Statistical Mechanics and its Applications 374.1 (2007): 457-470.

\bibitem{slowlearningmodel}
Norman, M. Frank. Markov processes and learning models. Vol. 84. New York: Academic Press, 1972.

\bibitem{JUNG}
O'Madadhain, Joshua et al. The JUNG (Java Universal Network/Graph) Framework No. UCI-ICS 03-17. (2003) 

\bibitem{disease}
Read, Jonathan M., Ken TD Eames, and W. John Edmunds. ``Dynamic social networks and the implications for the spread of infectious disease." Journal of the Royal Society Interface 5.26 (2008): 1001-1007.

\bibitem{MonteCarlo}
Rubinstein, Reuven Y., and Dirk P. Kroese. Simulation and the Monte Carlo method. Vol. 707. John Wiley \& Sons, 2011.

\bibitem{metric}
Schaeffer, Satu Elisa. ``Graph clustering." Computer Science Review 1.1 (2007): 27-64.

\bibitem{ncbi}
Vaquera, Elizabeth and Grace Kao (2008). ``Do You Like Me as Much as I Like You? Friendship Reciprocity and its Effects on School Outcomes among Adolescents." Social Science Research. 37: 55-72.

\bibitem{DataEx1}
Wen Dong, Bruno Lepri, and Alex Pentland. Modeling the co-evolution of behaviors and social relationships using mobile phone data. In Proceedings of the 10th International Conference on Mobile and Ubiquitous Multimedia (MUM), pages 134-143, 2011.

\end{thebibliography}
\end{document}